\documentclass[twocolumn,amssymb, nobibnotes, showpacs, superscriptaddress, aps, prd]{revtex4}
\usepackage{amsmath}
\usepackage{epsfig}
\usepackage{multirow}

\usepackage{graphicx}
\usepackage{dcolumn}
\usepackage{bm}

\usepackage[bookmarks=true,
   colorlinks=true,
   linkcolor=blue,
   urlcolor=blue,
   citecolor=blue,
   bookmarks=true,
   hyperindex=true
]{hyperref}

\begin{document}
\title{Extracting the differential phase in dual atom interferometers by modulating magnetic fields}
\author{Yu-Ping Wang}
\affiliation{State Key Laboratory of Magnetic Resonance and Atomic and Molecular Physics,
Wuhan Institute of Physics and Mathematics, Chinese Academy of Sciences, Wuhan
430071, China}
\affiliation{Center for Cold Atom Physics, Chinese Academy of Sciences, Wuhan 430071, China}
\affiliation{University of Chinese Academy of Sciences, Beijing 100049, China}
\author{Jia-Qi Zhong}
\author{Xi Chen}
\author{Run-Bing Li}
\email{rbli@wipm.ac.cn}
\affiliation{State Key Laboratory of Magnetic Resonance and Atomic and Molecular Physics,
Wuhan Institute of Physics and Mathematics, Chinese Academy of Sciences, Wuhan
430071, China}
\affiliation{Center for Cold Atom Physics, Chinese Academy of Sciences, Wuhan 430071, China}
\author{Da-Wei Li}
\author{Lei Zhu}
\author{Hong-Wei Song}
\affiliation{State Key Laboratory of Magnetic Resonance and Atomic and Molecular Physics,
Wuhan Institute of Physics and Mathematics, Chinese Academy of Sciences, Wuhan
430071, China}
\affiliation{Center for Cold Atom Physics, Chinese Academy of Sciences, Wuhan 430071, China}
\affiliation{University of Chinese Academy of Sciences, Beijing 100049, China}
\author{Jin Wang}
\email{wangjin@wipm.ac.cn}
\author{Ming-Sheng Zhan}
\affiliation{State Key Laboratory of Magnetic Resonance and Atomic and Molecular Physics,
Wuhan Institute of Physics and Mathematics, Chinese Academy of Sciences, Wuhan
430071, China}
\affiliation{Center for Cold Atom Physics, Chinese Academy of Sciences, Wuhan 430071, China}


\begin{abstract}
\textbf{Abstract}

We present a new scheme for measuring the differential phase in dual atom interferometers. The magnetic field is modulated in one interferometer, and the differential phase can be extracted without measuring the amplitude of the magnetic field by combining the ellipse and linear fitting methods. The gravity gradient measurements are discussed based on dual atom interferometers. Numerical simulation shows that the systematic error of the differential phase measurement is largely decreased when the duration of the magnetic field is symmetrically modulated. This combined fitting scheme has a high accuracy for measuring an arbitrary differential phase in dual atom interferometers.

\end{abstract}
\maketitle

\begin{flushleft}
\textbf{ 1. Introduction}
\end{flushleft}

Atom interferometers (AIs) have been successfully applied in many fields, such as determination of the Newtonian gravitational constant\cite{Fixler2007a,Lamporesi2008a,Rosi2014a}, test of the weak equivalence principle\cite{Schlippert2014a,Tarallo2014a,Zhou2015a}, and measurement of gravity\cite{Peters1999a,Peters2001a,Hu2013a}, gravity gradient\cite{McGuirk2002a,Sorrentino2012a,Duan2014a} and rotation\cite{Canuel2006a,Gauguet2009a,Stockton2011a,Berg2015a}. To suppress common phase noises, differential phase measurement techniques were developed in dual AIs. For example, vibrational noises can be suppressed in atom gravity gradiometers, and gravity and vibration induced common phase noises can be suppressed in atom gyroscopes by differential phase measurements. 

To accurately measure the differential phase of the dual AIs, various methods were used, including least-squares fitting\cite{McGuirk2001a}, ellipse fitting\cite{Foster2002a}, and Bayesian estimation\cite{Stockton2007a}. The least-squares fitting is an intuitive method, but it is only suitable for the case of the low common phase noise. The ellipse fitting can suppress the common phase noise, but its systematic error can not be ignored in small differential phase measurements. The differential phase of the dual AIs is usually very small in short baseline atom gravity gradiometers\cite{Wu2009a,Bertoldi2006a,Rakholia2014a,Biedermann2015a}. To measure the small differential phase by the ellipse fitting method, a bias magnetic field was applied to offset the phase difference of the dual AIs, where the magnetic field induced phase (MFIP) should be known as precise as possible. The MFIP was usually measured by Raman pulses\cite{Rakholia2014a} or extra microwave sequences\cite{Foster2002a}. However, uncertainty in the MFIP measurement usually causes systematic errors in differential phase measurements. The Bayesian estimation is more accurate, but it requires a priori phase noise model\cite{Rosi2015a}. The ellipse fitting and Bayesian estimation are modified to extract the differential phase with small systematic uncertainty\cite{Barrett2015a,Pereira2015a}.

In this paper, we present a combined fitting scheme for extracting the differential phase in dual AIs. The magnetic field is modulated in one atom interferometer, and the differential phase is extracted by using the ellipse and linear fitting methods. By the numerical simulation, the differential phases are extracted with the combined fitting scheme, and they are compared with the ellipse fitting method. The systematic error and standard deviation are analyzed when an arbitrary differential phase is extracted in dual AIs. This scheme is useful for accurately extracting the small differential phase, and the gravity gradient measurements are discussed based on dual AIs.

\begin{flushleft}
\textbf{2. Theoretical model}
\end{flushleft}

In the dual AIs, the atoms are simultaneously manipulated by common Raman lasers. After the atoms coherently interact with three Raman pulses, the transition probabilities, $P_{A}(k)$ for one interferometer and $P_{B}(k)$ for the other interferometer, are given by\cite{Foster2002a}
\begin{subequations}\label{eq1}
\begin{align}
& P_{A}(k)=\frac{1}{2}+\frac{1}{2}[1+c_{A1}(k)]\cos[\phi_{c}(k)+\phi_{d}+\phi_{A2}(k)],\\
& P_{B}(k)=\frac{1}{2}+\frac{1}{2}[1+c_{B1}(k)]\cos[\phi_{c}(k)],
\end{align}
\end{subequations}
where, $\phi_{c}(k)$ is the common phase, it contains the common phase noise. $\phi_{A2}(k)$ is the differential phase noise. $c_{A1}(k)$ and $c_{B1}(k)$ are the amplitude noises of the dual AIs. $P_{A}(k)$ and $P_{B}(k)$ are the sinusoidal signals, which can be observed by scanning $\phi_{c}(k)$ from 0 to $2\pi$. When a modulated magnetic field is applied in one of the dual AIs, the differential phase of two sinusoidal signals is
\begin{equation}\label{eq2}
\phi_{d}=\phi_{d0}+\phi_{m},
\end{equation}
where, $\phi_{d}$ and $\phi_{d0}$ are the differential phase with and without modulating the magnetic field, respectively. $\phi_{m}$ is the MFIP with the magnetic field modulated. In the dual AIs, one AI experiences an uniform magnetic field $B_{0}$ while the other AI undergoes a hybrid magnetic field composed of $B_{0}$ and $B(t)$. $B(t)$ is a square wave magnetic field pulse with the amplitude $B$ and the duration $t_{m}$. Therefore, the phase shift $\phi_{m}$, caused by the hybrid magnetic field, is given by\cite{Gouet2008a}

\begin{eqnarray}
\phi_{m}&=&2\pi K\int^{T}_{-T} g(t)[B(t)+B_{0}]^2 dt \nonumber\\
&=&2\pi K\int^{T}_{-T} g(t)B_{0}^2 dt \nonumber\\
& &+2\pi K\int^{-t+t_{m}}_{-t}g(t)[2BB_{0}+B^2]dt,
\end{eqnarray}
where, $K$ is the coefficient of the quadratic Zeeman shift, $g(t)$ is the sensitivity function, and $T$ is the time interval between two consecutive Raman pulses. The constant uniform magnetic field doesn't induce extra phase shift, because $g(t)$ is an odd function in the range of $[-T, T]$. Thus, the integral of $2\pi K\int^{T}_{-T} g(t)B_{0}^2 dt$ is zero, and Eq.\,(3) is written as
\begin{equation}\label{eq4}
\phi_{m}=\alpha t_{m},
\end{equation}
where, $\alpha=2\pi K[2BB_{0}+B^2]$, and $\phi_{m}$ can be linearly controlled by $t_{m}$. When $\phi_{d0}$ is small, $\phi_{d}$ can be set to $\pi/2$ by modulating $t_{m}$. For an arbitrary $\phi_{d0}$, an ellipse can be obtained by shifting $\phi_{d}$ to $(2M+1)\pi/2$, where $M$ is an integer number. For a given $t_{m}$, $\phi_{d}$ can be extracted by the ellipse fitting method. For the different $t_{m}$, $\phi_{d}$ can also be obtained by the ellipse fitting. For the $i$-$th$ modulated magnetic field, the differential phase is expressed as
\begin{equation}\label{eq4}
\phi_{d}(i)=\phi_{d0}+\alpha t_{m}(i),
\end{equation}
where, $\phi_{d}(i)$ is the differential phase extracted by the ellipse fitting method, $\alpha t_{m}(i)$ is the MFIP, and $t_{m}(i)$ is the duration of the magnetic field. According to Eq.(5), $\phi_{d}$ is a linear function of $t_{m}$, and $\phi_{d0}$ can be extracted from the intercept of the fitted line.
Although the magnetic field is modulated, the differential phase can be directly extracted without measuring the MFIPs. This means that the systematic error can be reduced with the combined fitting scheme.

\begin{flushleft}
\textbf{3. Numerical simulation}
\end{flushleft}

\begin{figure}[htp]
	\centering
		\includegraphics[width=0.50\textwidth]{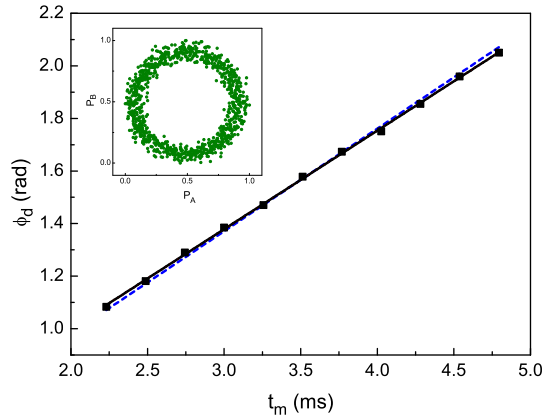}
	\caption{(Color online) Dependence of the differential phase on the magnetic field duration when $\phi_{d}$ is distributed around $\pi/2$. $\phi_{d}$ is extracted by the ellipse fitting (black squares and solid line), and it is also calculated by the same parameters (blue dashed line). $\phi_{d0}$ is extracted by the intercept of the fitted line (black solid line).}%
	\label{fig-1}%
\end{figure}

We numerically simulate the differential phase extraction under the conditions of $\phi_{d0}=0.20$ rad, $B_{0}=20$ mG, $B=200$ mG, and $K=1.29$ kHz/G$^2$ for $^{85}$Rb\cite{Li2009a,Zhou2011a}. The normally distributed random noise is considered in Eqs.\,(1a) and \,(1b) with the common phase noise variance of 0.25 in $\phi_{c}(k)$, the differential phase noise variance of 0.01 in $\phi_{A2}(k)$, and the amplitude noise variances of 0.01 in $c_{A1}(k)$ and $c_{B1}(k)$. For $t_{m}=3.5$ ms, $P_{A}(k)$ and $P_{B}(k)$ are numerically simulated by scanning $\phi_{c}(k)$ from 0 to $2\pi$ with 100 steps, which are plotted as a typical ellipse after running 10 cycles as shown in Fig.1 (insert). For the different $t_{m}$, a series of ellipses are created by the same way, and the corresponding differential phases are extracted by the ellipse fitting, as shown in Fig.1 (black solid squares). $\phi_{d0}$ is extracted from the intercept of the fitted line (black solid line), and it is ($0.25 \pm 0.01$) rad. According to Eq.(5), the theoretical values of $\phi_{d}$ for the same parameters are shown in Fig.1 (blue dashed line). $\phi_{d}$ is consistent with the theoretical value ($\pi/2$) when $t_{m}$ is 3.5 ms. However, $\phi_{d}$ is higher than the theoretical value (overestimated) when $t_{m}$ is shorter than 3.5 ms, while it is lower than the theoretical value (underestimated) when $t_{m}$ is longer than 3.5 ms. There is an error of 25\% between the fitted value and theoretical value, which is caused by the overestimated and underestimated values of the ellipse fitting.

\begin{figure}[htp]
	\centering
		\includegraphics[width=0.50\textwidth]{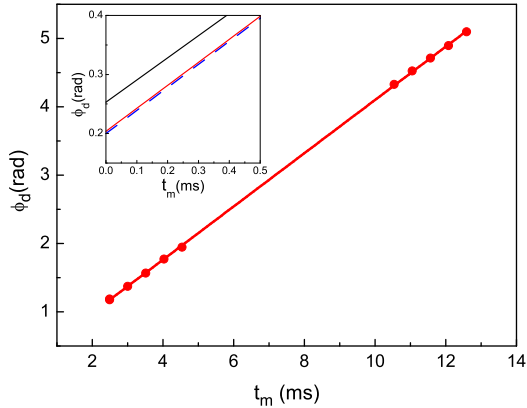}
	\caption{(Color online) Dependence of the differential phase on the magnetic field duration when $\phi_{d}$ is symmetrically distributed around $\pi/2$ and $3\pi/2$. $\phi_{d}$ is extracted by the ellipse fitting (red circles), and $\phi_{d0}$ is extracted by the intercept of the fitted line (red solid line). The insert shows that the fitted results (red and black solid lines) and the theoretical value (blue dashed line).}%
	\label{fig-1}%
\end{figure}

We noticed that the overestimated and underestimated values of the ellipse fitting are symmetrically distributed with respect to the theoretical values except for $(2M+1)\pi/2$. Therefore, the combined fitting scheme can be optimized by shifting $\phi_{d}$ to $\pi/2$ and $3\pi/2$. Similar to Fig.1, the dependence of $\phi_{d}$ on $t_{m}$ is simulated by the same parameters, as shown in Fig.2 (red solid circles). $\phi_{d0}$ is also extracted from the intercept of the fitted line (red solid line). The extracted differential phase of ($0.20\pm0.01$) rad agrees very well with the given value (0.20 rad). The fitted lines in Fig.1 (black solid line) and Fig.2 (red solid line) are compared with the theoretical line (blue dashed line), as shown in Fig.2 (insert). When $\phi_{d}$ is symmetrically chosen around $\pi/2$ and $3\pi/2$, the fitted error of the differential phase can be completely ignored, because the overestimated and underestimated errors are cancelled. This implies that the fitted error of the differential phase measurement is caused by the ellipse fitting rather than the linear fitting.

\begin{flushleft}
\textbf{4. Performance evaluation}
\end{flushleft}

To evaluate the performance of the combined fitting scheme, we simulate the systematic error and standard deviation of an arbitrary differential phase $\phi_{d0}$ in the range of [0, $\pi$]. For the $j$-$th$ differential phase measurement, the systematic error $\sigma_{sys}(j)$ and the standard deviation $\sigma_{std}(j)$ are defined as
\begin{equation}\label{eq1b}
	\sigma_{sys}(j)=E(\hat{\phi_{d0}}(j)-\phi_{d0}(j)),
\end{equation}
and
\begin{equation}\label{eq1b}
\sigma_{std}(j)=\sqrt{E(\hat{\phi_{d0}}(j)-E(\hat{\phi_{d0}}(j)))^2}.
\end{equation}

The total error $\sigma(j)$ for the $j$-$th$ case is given by
\begin{equation}\label{eq1b}
\sigma(j)=\sqrt{\sigma_{sys}(j)^2+\sigma_{std}(j)^2},
\end{equation}
where, $E(x)$ represents the ensemble average of a random variable $x$. $\hat{\phi_{d0}}(j)$ is the fitted value, and $\phi_{d0}(j)$ is the real value for the $j$-$th$ case, respectively.

\begin{figure}[htp]
	\centering
		\includegraphics[width=0.50\textwidth]{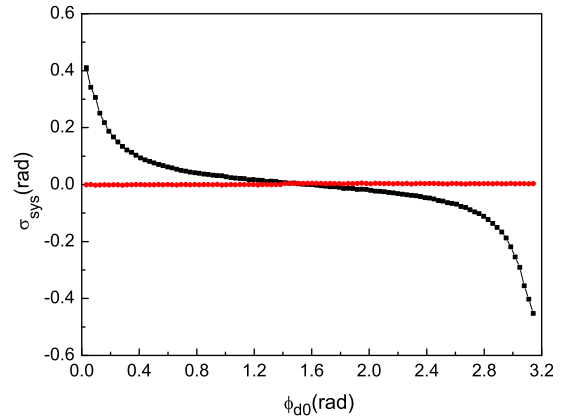}
	\caption{(Color online) Dependence of the systematic error on the differential phase. They are calculated by using the ellipse fitting method (black solid squares) and the combined fitting scheme (red solid dots).}%
	\label{fig-1}%
\end{figure}

The systematic error and standard deviation are numerically calculated for each differential phase in the range of [0, $\pi$]. Each $\hat{\phi_{d0}}$ is simulated by 100 cycles when the random distributed noise is considered, where $\hat{\phi_{d0}}$ is extracted similar to Fig.2. From Eqs.(6) and (7), the systematic errors and standard deviations can be calculated. According to Eq.(6), the dependence of the systematic error on the differential phase is shown in Fig.3. The black solid squares are the systemic errors based on the ellipse fitting method, where the systematic errors are very bad around 0 and $\pi$. The systematic error is positive for $\phi_{d0}<\pi/2$, while it is negative for $\phi_{d0}>\pi/2$. The red solid circles are the systemic errors based on the combined fitting scheme, where $\hat{\phi_{d0}}$ is obtained by shifting $\phi_{d0}$ to $\pi/2$ and $3\pi/2$ for $\phi_{d0}<1.4$ rad, while it is obtained by shifting $\phi_{d0}$ to $3\pi/2$ and $5\pi/2$ for $\phi_{d0}>1.4$ rad. The systematic errors are very small in the whole range from 0 to $\pi$, which implies the combined fitting scheme is better than the ellipse fitting method. According to Eq.(7), the dependence of the standard deviation on the differential phase is shown in Fig.4, where the data of $\hat{\phi_{d0}}$ and $\phi_{d0}$ are same as in Fig.3. The black solid squares are the standard deviations based on the ellipse fitting method, while the red solid circles are the standard deviations based on the combined fitting scheme. When $\phi_{d0}$ is around 0 and $\pi$, the standard deviations are very large with the ellipse fitting method, while they are very small with the combined fitting scheme. This implies that the standard deviations with the combined fitting scheme are also better than those with the ellipse fitting method. The systematic errors are very small with the combined fitting scheme as in Fig.3 (red solid circles), but the standard deviations are limited by the amplitude noise and differential phase noise as in Fig.4. Nevertheless, the standard deviations is largely reduced with the combined fitting scheme, especially for the small differential phase measurement.

\begin{figure}[htp]
	\centering
		\includegraphics[width=0.50\textwidth]{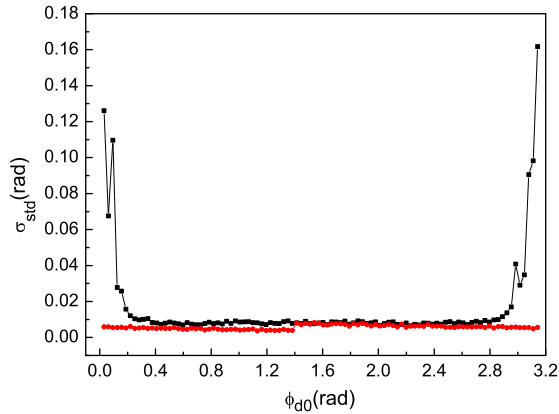}
	\caption{(Color online) Dependence of the standard deviation on the differential phase. They are calculated by using the ellipse fitting method (black solid squares) and the combined fitting scheme (red solid dots).}%
	\label{fig-1}%
\end{figure}

To discuss the accuracy of an atom gravity gradiometer, we consider two $^{85}$Rb AIs with the baseline of 1 m and the time interval of 100 ms. The corresponding differential phase is 240 mrad for the horizontal gravity gradient (HGG) of 1500 E, and it is 480 mrad for the vertical gravity gradient (VGG) of 3000 E on the surface of the Earth. The total errors are calculated by the numerically simulation, as shown in Table I. For the HGG, the standard deviation and systematic error are 12.6 mrad and 154.8 mrad with the ellipse fitting method, while they are only 4.9 mrad and 0.9 mrad with the combined fitting scheme. The total error with the combined fitting scheme is suppressed by 31 times than that with the ellipse fitting method. For the VGG, the standard deviation is suppressed from 8.0 mrad with the ellipse fitting method to 4.1 mrad with the combined fitting scheme, and the systematic error is reduced from 79.8 mrad to 0.3 mrad. The total error is suppressed by 21 times. Therefore, the combined fitting scheme is better than the ellipse fitting method in processing the systematic error and the standard deviation. This implies that the accuracy can be significantly improved in gravity gradient measurements.

\vskip 10pt
\noindent

\noindent{\footnotesize Table I. Comparison between the ellipse fitting method and the combined fitting scheme (unit: mrad).

\vskip 5pt
\noindent

\begin{tabular}{ccccccc}\hline
\multirow{2}{*}{Type} & \multirow{2}{*}{$\phi_{d0}$} & \multicolumn{2}{c}{Ellipse fitting method} &\multirow{2}{*}{}& \multicolumn{2}{c}{Combined fitting scheme} \\

\cline{3-4} \cline{6-7}

& $ $ & $~~~\sigma_{std}$ & $\sigma_{sys}$ &&$~~~~~~\sigma_{std}$ & $\sigma_{sys}$ \\
\hline
$HGG$ & $~~240~~$ & $~~~\pm 12.6$ & $+154.8$ &&$~~~~\pm 4.9$ & $+0.9$ \\
$VGG$ & $~~480~~$ & $~~~\pm 8.0$ & $+79.8$ &&$~~~~\pm 4.1$ & $+0.3$ \\
\hline
\end{tabular}}

\begin{flushleft}
\textbf{5. Conclusion}
\end{flushleft}

In summary, we proposed a combined fitting scheme for extracting an arbitrary differential phase in the dual AIs. The differential phase can be extracted by modulating the magnetic field. Although the extra magnetic field is applied, its amplitude need not be measured. This scheme avoids the disadvantages of the ellipse fitting method for extracting the small differential phase, and it is useful for measuring the arbitrary differential phase of the dual AIs. The numerical simulation implies that the systematic error and standard deviation with the combined fitting scheme are better than those with the ellipse fitting method, and the accuracy of the gravity gradient measurement can be significantly improved.

\begin{flushleft}
\textbf{Acknowledgment}
\end{flushleft}

We acknowledge the financial support from the National Natural Science Foundation of China under Grant Nos. 11227083, 91536221.

\end{document}